\documentclass[sigconf]{acmart}

\AtBeginDocument{%
  \providecommand\BibTeX{{%
    \normalfont B\kern-0.5em{\scshape i\kern-0.25em b}\kern-0.8em\TeX}}}

\usepackage{optidef}

\copyrightyear{2024}
\acmYear{2024}
\setcopyright{acmlicensed}\acmConference[KDD '24]{Proceedings of the 30th ACM SIGKDD Conference on Knowledge Discovery and Data Mining}{August 25--29, 2024}{Barcelona, Spain}
\acmBooktitle{Proceedings of the 30th ACM SIGKDD Conference on Knowledge Discovery and Data Mining (KDD '24), August 25--29, 2024, Barcelona, Spain}
\acmDOI{10.1145/3637528.3671597}
\acmISBN{979-8-4007-0490-1/24/08}

\begin{document}

\title{Multi-objective Learning to Rank by Model Distillation}

\author{Jie Tang}
\orcid{0009-0006-9509-2367}
\affiliation{%
  \institution{Airbnb}
  \city{San Francisco}
  \country{USA}
}
\email{jie.tang@airbnb.com}

\author{Huiji Gao}
\orcid{0009-0006-0424-248X}
\affiliation{%
  \institution{Airbnb}
  \city{San Francisco}
  \country{USA}}
\email{huiji.gao@airbnb.com}

\author{Liwei He}
\orcid{0009-0007-2942-3145}
\affiliation{%
  \institution{Airbnb}
  \city{San Francisco}
  \country{USA}}
\email{liwei.he@airbnb.com}

\author{Sanjeev Katariya}
\orcid{0009-0008-1519-174X}
\affiliation{%
 \institution{Airbnb}
 \city{San Francisco}
 \country{USA}}
\email{sanjeev.katariya@airbnb.com}

\renewcommand{\shortauthors}{Jie Tang, Huiji Gao, Liwei He, and Sanjeev Katariya}

\begin{abstract}
  In online marketplaces, search ranking's objective is not only to purchase or conversion (primary objective), but to also the purchase outcomes(secondary objectives), e.g. order cancellation(or return), review rating, customer service inquiries, platform long term growth. Multi-objective learning to rank has been widely studied to balance primary and secondary objectives. But traditional approaches in industry face some challenges including expensive parameter tuning leads to sub-optimal solution, suffering from imbalanced data sparsity issue, and being not compatible with ad-hoc objective. In this paper, we propose a distillation-based ranking solution for multi-objective ranking, which optimizes the end-to-end ranking system at Airbnb across multiple ranking models on different objectives along with various considerations to optimize training and serving efficiency to meet industry standards. We found it performs much better than traditional approaches, it doesn't only significantly increases primary objective by a large margin but also meet secondary objectives constraints and improve model stability. We also demonstrated the proposed system could be further simplified by model self-distillation. Besides this, we did additional simulations to show that this approach could also help us efficiently inject ad-hoc non-differentiable business objective into the ranking system while enabling us to balance our optimization objectives.
\end{abstract}



\settopmatter{printacmref=false}
\setcopyright{none}
\renewcommand\footnotetextcopyrightpermission[1]{}
\pagestyle{plain}
\maketitle

\section{Introduction} \label{intro}
Online marketplaces such as Amazon, eBay, Walmart, Airbnb, Doordash, Uber provide search functions for users to find their preferred product items. While these platforms vary in the types of services or goods they offer, from retail products to food delivery to rentals, they share similar challenges in their search ranking algorithms. In general, aiming for higher conversion rates is insufficient, as it is crucial to consider the outcomes of those conversions, such as order cancellations or returns, customer service inquiries, review ratings, and the long-term growth of the marketplace. Although a higher conversion rate can increase marketplace revenues, the associated costs of cancellations, returns, and customer service can significantly eat into profits. Additionally, an overemphasis on short-term conversion gains can jeopardize long-term growth by favoring established items over new listings, thus limiting the opportunities for new products to be discovered.\par
 To efficiently balance multi-objectives in search ranking, multi-objective optimization technologies have been recently studied for search ranking. For example, \cite{Yong22} reviewed multi-objective recommender systems and ranking algorithms, it classified most of previous work into two categories: scalarization methods (e.g. \cite{Xiao19}) and multi-objective evolutionary algorithms (MOEAs) (e.g.\cite{Marco12}).  In industrial practice, scalarization is much more popular than MOEAs given it's simple and also easy to be scaled to large training data efficiently. This paper's proposed approach is also connected to scalarization techniques, and thus, we will concentrate on comparing it with current scalarization methods. \par
    The most popular and intuitive approach in industry is model fusion  \cite{David20}, which trains one model for each objective independently, then combines them by weighted sum. This approach is simple but the drawback is also obvious: since objectives could interfere each other(the model which is optimized for one objective could hurt another one), simply combining them would lead  to a sub-optimal solution. Therefore a better approach is assigning weight to each objective's cost function, then sum them together as a single objective cost function, such that the multi-objective optimization(MOO) problem is converted to single objective optimization problem\cite{Kaisa98}. This approach is also called scalarization method, it's widely adopted in industry\cite{Xiao19} \cite{Ruobing21} \cite{Guy18} \cite{Yulong20}  \cite{ChunHow23}. Another related method is \(\epsilon\)-Constraint method\cite{Debabrata23} which takes one primary objective as optimization goal while secondary objectives as constraints. Besides optimizing aggregated objective cost functions, some study aggregated multiple labels into one label thus it could also construct a single objective cost function from multi-objective cost functions\cite{David20}, while as \cite{Debabrata23} pointed out, it could still be considered as a special type of scalarization method. \par
Those methods were discussed a lot in past regarding how to do better optimization, how to find better trade-off among optimization solutions. But one thing people may ignore yet important is imbalanced data across different objectives, e.g. there could be much more training data for clicks and conversions, but much less data for order cancellation and customer service inquiries etc due to event frequency. The scalarization approach is often implemented as multi-task learning scheme  \cite{ChunHow23} \cite{Ozan18} \cite{Zhe18}, one advantage of multi-task learning is the shared bottom layers could learn shared representation across all tasks(objectives), this way task(objective) with less data could benefit from task(objective) with more data. While this advantage could also harm learning when correlation among tasks is low, this could especially be true in marketplace search ranking, indeed objectives are often in opposite directions : focusing more on long term growth may hurt conversions (which are usually measured in short term), reducing customer service inquiries may hurt both conversions and long term growth since model may rank well-established items with higher quality/price to top. Thus with more and more conflicted objectives are added into multi-task model, it's expected to observe less efficient learning and sub-optimal solution. \par
    Though multi-task learning approach could be sub-optimal when objectives are not highly correlated, it still performs much better than simple model fusion thanks to the power of shared representation.  This is also verified in Airbnb practice\cite{ChunHow23},  the multi-task learning system implemented in Airbnb co-trains models for all objectives and achieved significant business metrics improvements. One learning from this system is tuning two sets of parameters could be very challenging: one for training loss weights which combines objectives' loss functions, another one for online score aggregation weights at serving stage. The second one is usually determined in ad-hoc way and sometimes it asks for online grid search which is cost-inefficient, even so the found weights still can't be guaranteed to be optimal. Besides this, we found the online score aggregation could be unstable, the weights tuned for one set of models may not work well with another set of models. Thus the model performance would be degraded over time or expensive tuning has to be done each time when any model is updated. The instability is not only from ad-hoc weight tuning, in our experiments it shows even simply retraining could also cause not small metric change, such that it may break balance which was optimized in previous model for different objectives. \par
    Besides the the challenges mentioned above, we also found another issue ignored by previous studies: nowadays most of ranking systems are deep learning models, the scalarization method assumes each objective cost function is differentiable, therefore the aggregated single objective cost function could be trained by backpropagation. While in practice, this is not always true, sometimes the objective could even be just an ad-hoc rule : e.g. show reasonable percentage of new items thus it would help long term growth. In this case, such objective can't be included into optimization directly, it has to be some manual tuning after the model is trained and deployed, though such manual tuning would degrade model performance.\par
    Recently, another trend is to apply model distillation into ranking and recommender systems \cite{Jiaxi18} \cite{Honglei21} \cite{Franco23}, \cite{Honguk22}, almost all existing methods in this domain focus on two topics : 1) distillate large and complicated model into compact and efficient model; 2)distillate ensemble models into one single model. The main purpose is to save inference cost while retain model performance. Upon observing that the loss functions of model distillation and the scalarization method for multi-objective optimization share a similar structure, we realized the integration of model distillation and the multi-objective learning to rank could help address challenges discussed above. Therefore in this paper, we describe a general model distillation approach for optimizing multi-objective in search ranking and recommender system. To our best knowledge, this is the first attempt to apply model distillation in multi-objective learning to rank. Our major contributions include:
    \begin{itemize}
        \item We reformulate the multi-objective learning to rank problem (MO-LTR) as a model distillation problem which could mitigate imbalanced data issue, get rid of online score aggregation weights tuning, therefore we achieve better primary objective while meet secondary objectives constraints.
        \item We extend the distillation based multi-objective ranking algorithm by introducing soft-label concept into MO-LTR and demonstrate it could help reduce model irreproducibility and simplify proposed ranking system by self-distilling soft-labels.
        \item We also show that ad-hoc non-differentiable business objectives could be easily injected into ranking model by revising soft-labels, thus non-differentiable objective could be easily included  into MO-LTR.
    \end{itemize}

\section{RELATED WORK} \label{related_work}
 Learning to rank(LTR) has been a popular research topic for decades, one important branch of previous studies is to evolve loss schemes: 
 \begin{itemize}
     \item The point-wise loss predicts action probability (e.g. pCTR, pCVR) for each item separately\cite{McMahan13}.
     \item The pair-wise loss looks at two items each time, and converts ranking to a binary classification problem : whether item A is better than item B. \cite{Burges05} \cite{burges2010from}
     \item The list-wise loss considers the whole list of items and try to approximate the optimal order. \cite{cao07}
 \end{itemize}
In early days, these studies only considered optimizing single objective: NDCG, since the major LTR application that time was web search. Recently, with emergence of online marketplace, search and recommender system become very popular in this new domain. Unlike web search,   online marketplace is a two-sided market, both user journey and merchant journey is much longer than web search user: a typical online marketplace user would do comparison shopping to purchase one item by browsing lots of items,  then later the purchased item could still be cancelled or returned by user; if user is not satisfied with the purchase, he or she may complain to customer service. Accordingly merchant would also go through the similar journey. In this case, only optimizing for NDCG is not enough,  industrial companies start applying multi-objective optimization(MOO) into learning-to-rank: \cite{Mario12} applied MOO to talent match system, \cite{David20} applied MOO to balance two objectives revenue and purchase, \cite{Michinari19} optimized multi-objectives including relevance, purchase, quality, rating, return, etc, \cite{Xiao19} proposed a pareto-eficient algorithm to balance GMV and CTR, \cite{ChunHow23} optimized Airbnb search journey with multi-task learning by considering objectives including click, booking, cancellation, rejection, \cite{Zhishan23} optimized for pre-ranking/ranking consistency, etc. Most of these researches applied scalarization method which converts multi-objective optimization to single-objective optimization:\par

\[min \sum_{k} \omega_k \cdot C_k(X)\tag{1} \label{1}\]
where \(\omega_k\) is assigned weight to \(k^{th}\) objective, it represents priority or importance of \(k^{th}\) objective, and could be assigned manually or adaptively, \(C_k(x)\) is the cost function of \(k^{th}\) objective, \(X\) is training data. Similar to \cite{David20}, we could further rewrite \(C_k(X)\) as loss function

\[C_k(X)=\sum_{i} loss(f(\theta_s, \theta_k, X_{i}), L_{i,k})\tag{2} \label{2}\]

where \(f(\theta_s, \theta_k, X_i)\) is the ML model for \(k^{th}\) objective, \(\theta_s\) is shared model parameters across all objectives, \(\theta_k\) is model parameters specifically for \(k^{th}\) objective, \textit{\ \(\{X_{i}, L_{i,k}\}\)}  is feature vector and label of each training example for \(k^{th}\) objective. With this setup, there could be following 4 variants:
\begin{enumerate}
    \item \(\theta_s\) = \(\varnothing\),  \(\theta_k \neq \varnothing\) 
    \begin{itemize}
        \item  models don't share any parameters, but they are jointly trained with an aggregated loss function. At serving time, final ranking score is aggregated from all models' scores.
    \end{itemize}
    \item   \(\theta_s\neq\varnothing\),  \(\theta_k \neq \varnothing\) \label{multi_task}
    \begin{itemize}
        \item This is a typical multi-task learning setup for deep learning model, models share bottom layers to learn shared representation. At serving time, final ranking score is aggregated from all models' scores. e.g.  \cite{ChunHow23} \cite{Ozan18}
    \end{itemize}
    \item \(\theta_s\neq\varnothing\),  \(\theta_k = \varnothing\) \label{variant_3}
    \begin{itemize}
        \item With this setup, there is only one single model to fit multi-objectives. At serving time, the model score could be directly used as ranking score.  e.g. \cite{David20} \cite{Xiao19}
    \end{itemize}
    \item \(\theta_s=\varnothing\),  \(\theta_k = \varnothing\)\label{model_fusion}
    \begin{itemize}
        \item There is no trainable parameter for each model, this means each model is pre-trained  to optimize  \(C_i(x_i)\) separately. At serving time, final ranking score is aggregated from all models' scores.
    \end{itemize}
\end{enumerate}
\par
Both Variants 1) and 2) are multiple-task learning setup, while 2) is more popular since it's expected that the shared bottom layer could learn shared and better representation, and also help data sparsity issue. There are two sets of weights need to be tuned in 1) and 2): one for \(\omega_i\)  in training loss \eqref{1}, another one for scores aggregation weights at model serving stage. Advantage of variant 3) is score from the single model could be directly used at serving time, while training one single model to fit multi-objective is more challenging.  Variant 4) is actually model fusion, it's also popular due to its simplicity, but it may perform worse.\par
To our best knowledge,  the most relevant previous work to ours is \cite{David20},  which proposed two approaches:
\begin{enumerate}
    \item Stochastic label aggregation: For each training example, the label is randomly sampled from a label set, each label in this label set is mapped to one objective.
    \item Two-phase model combination: At first step, each model is trained to optimize different objectives separately. At second step, the model scores are features of another model which would be trained with  stochastic label aggregation approach.

\end{enumerate}
\par
Though \cite{David20} shows some theoretical advantage of their approaches, as we pointed earlier it may not be efficient with extremely imbalanced training data across objectives, since with the single model setup, minor objective may be overwhelmed by objective with much larger data. In Airbnb ranking system, training data among objectives are highly imbalanced, in worst cast, the label imbalance ratio could be more than 10, while the training data in \cite{David20} is much more balanced: 2 datasets are well balanced, another one's unbalance ratio is 3.  Also all 4 scalarization variants mentioned above can't handle non-differentiable objective e.g. manual rules which are usually applied in industrial world.\par

\section{Proposed framework}\label{framework}

\subsection{Problem Statement} \label{problem_statement}

Given a search query  \(q_i\) , there could be \(n\) matched items(returned from retrieval stage), the goal of learning to rank is to assign a score \(s^j\) to \(j^{th}\) item  \(t_{i,j}\)  so that those items could be ranked in descending order of scores. In ML terminology, here the training example is \(\{q_i, X_i, L_i\}\), where \(X_i=\{x_{i,j}\}_{j=1}^n\), \textit{\(x_{i,j}\in R^{m\times1}\)} is feature vector of \(t_{i,j}\).  \(L_i = \{{l_{i,j}}\}_{j=1}^n\), \textit{\(l_{i,j}\in R^{K\times1}\)} is  label vector of \(t_{i,j}\), \(K\) is number of objectives. Ideally for each item, it's expected there is one label for each objective, but due to data sparsity, the label could be missing,  thus though it's assumed all objectives share the same set of training data, some objective could only be trained over a small subset of training data when it's trained separately.\par
For each objective, its cost function is defined like \eqref{2} 

\[C_k(X)=\sum_{i} loss(f(\theta, X_{i}), L_{i,k})\label{2.1}\tag{2.1}\]
where \(\theta\) is trainable model parameters. Since \(X_i\) contains \(n\) items, the \(loss(\cdot)\) function here is Listwise Loss\cite{cao07}.   \(f(\theta, X) \in R^{n\times1}\) is prediction of the model which optimize those \(K\) objectives' cost function at the same time and also find better balance among them, \(f(\theta, X_i)\mid_p\) is denoted as prediction of \(p^{th}\) item in \(X_i\)\par

\subsection{From Multi-Objective Optimization To Model Distillation}\label{softlabel}
In practical operations, major online marketplaces usually takes one business objective(e.g. CVR) as primary objective, while considering other secondary business objectives as constraints(e.g. cancellations, returns, review ratings). This is more aligned with \(\epsilon\)-Constraint method\cite{Debabrata23} \cite{Yong22}, it could be written as 

\begin{mini*}|s|
{}{C_1(X)}
{}{}
\addConstraint{C_k(X) \leq\varepsilon_k , k =  2, ..., K  }\tag{$3$} \label{3}
\end{mini*}

where \(C_k(X)\) is \(k^{th}\) objective's cost function, here \(C_1(X)\) is set as primary objective cost function, \(\varepsilon_k\) is upper bound cost of each secondary objective.  \par
 If each objective cost function is optimized separately, for each objective a model \(f_k(\theta^{*}_k,X)\) could be trained without considering other objectives by only optimizing \(C_k(X)\). The corresponding cost \(C^{*}_k(x)\) could be considered as the lower bound of  \(C_k(X)\), since a model optimizes for multiple objectives can't perform better than a dedicated model. This way we could rewrite each constraint as 

\[ C^{*}_k(X) \leq C_k(X)  \leq C^{*}_k(X) + \varepsilon^{'}_k \Rightarrow \mid C_k(X) - C^{*}_k(X) \mid \leq \varepsilon^{'}_k\tag{4} \label{4} \]

 This means we could tolerate \(k^{th}\) objective \(C_k(X)\) performs at most \(\varepsilon^{'}_k\) worse than  \(C^{*}_k(X)\).  \par
 \(C_k(X)\) could be rewritten with \eqref{2.1} as function of the underline model \(f(\theta,X)\)
\[
C_k(X)=\sum_{i} loss(f(\theta, X_{i}), L_{i,k})=C_k(f(\theta,X))\tag{5} \]
\[
C_k^*(X)=C_k(f_k(\theta^{*}_k,X))\tag{6} \]
 where \(f(\theta,X)\) is the single model tries to optimize multi-objectives \textit{\(\{C_k(X)\}\)} ,  \(f_k(\theta^*_k,X)\) is the model trained for optimizing \(C_k(X)\) only.  \par
If each objective cost function is Lipschitz continuous, we have 
 
\[\mid C_k(f(\theta, X)) - C_k(f_k(\theta^*_k,X))\mid \leq M\mid f(\theta, X) - f_k(\theta^*_k,X)\mid \label{7} \tag{7}\]

where \(M\) is a constant called as Lipschitz constant.\par
Combine \eqref{4} and \eqref{7}, we could find a surrogate constraint
\[\mid f(\theta,x) - f_k(\theta^*_k,X)\mid \leq\varepsilon^{''}_k \label{8} \tag{8}\]
\par
So now we could rewrite \eqref{3} as 

\begin{mini*}|s|
{\theta}{C_1(f(\theta,X))}
{}{}
\addConstraint{ \mid f(\theta,X) - f_k(\theta^*_k,X)\mid \leq\varepsilon^{''}_k , k =  2, ..., K  }\tag{$9$} \label{9}
\end{mini*}

\par
This optimization problem optimizes primary objective with a model \(f(\theta,x)\) which also approximates each objective's own optimization solution \(f_k(\theta^*_k,X)\) at the same time.  With Lagrangian relaxation, we could further rewrite \eqref{9} as

\[\underset{\theta}{\min}\; C_1(f(\theta,X)) + \sum_{k=2}^{K} (\omega_k\mid f(\theta,X)-f_k(\theta^*_k,X) \mid )\tag{10} \label{10} \]

Here Euclidean distance could be replaced with KL distance, and in LTR context,  cross entropy (which is Lipschitz continuous) is usually adopted as loss function for each objective. Since KL distance is equivalent to cross entropy,  \eqref{10} becomes

\[\underset{\theta}{\min} \; CE(f(\theta,X), L_1) + \sum_{k=2}^{K} \omega_k CE(f(\theta,X), f_k(\theta^*_k,X)) \tag{11} \label{11} \]
where \(CE(\cdot)\) is the cross-entropy loss,  \eqref{11} could be further simplified as (short proof could be found in  \nameref{proof})
\[\underset{\theta}{\min} \; CE(f(\theta,X), L_1) +  CE( f(\theta,X), \sum_{k=2}^{K}\omega_k  f_k(\theta^*_k,X))\tag{12} \label{12} \]
\par
In \eqref{9}, it also makes sense to approximate primary objective's own optimization solution as additional constraint, thus we have 
\[\underset{\theta}{\min} \; CE(f(\theta,X), L_1) +  CE(f(\theta,X), \sum_{k=1}^{K}\omega_k f_k(\theta^*_k,X))\tag{13} \label{13} \]
Now we have a loss(single objective) function which is very similar to model distillation loss function  (e.g. \cite{Geoffrey14}   \cite{Honglei21}): the first term is a loss between the prediction and primary objective's ground truth label (it's also referred as "hard label" in this paper), the second term is a distillation loss pushes model to approximate a soft-label which is computed by aggregating each objective's own optimization solution. From \eqref{13}, this soft-label is defined as
\[\sum_{k=1}^{K}\omega_k f_k(\theta^*_k,X)\label{14}\tag{14}\]

which could also be considered as model fusion of each objective's own optimization solution. Thus  \eqref{13} shows model fusion and model distillation could be combined together for multi-objective optimization. \par
So to optimize multi-objectives in context of learning-to-rank, this formula tells us that we need to
\begin{enumerate}
    \item Train a model \(f_k(\theta^*_k,X)\) for each objective separately without considering other objectives. This is done by training model with each objective's label separately.
    \item For each training example, soft-label is generated as\\  \(\sum_{k=1}^{K}\omega_k f_k(\theta^*_k,X)\)
    \item Run model distillation by considering both soft-labels and primary objective's ground-truth labels(hard labels), then we get a model which optimizes primary objective under secondary objectives' constraints. \par
\end{enumerate}

\section{Multi-objective learning to rank system}\label{system} 
\subsection{System Overview}
To design a multi-objective learning to rank system as Eq \eqref{13} described, at first we rewrite Eq\eqref{13} by introducing an extra weight \(\alpha\) to balance learning from hard-labels and soft-labels, since as experiment in section \ref{exp} showed assigning equal weights to hard-label and soft-label is not optimal in practice. \par
Thus by also expanding  \(CE(\cdot)\), the loss function \eqref{13} is rewritten as 
\[Loss = -\alpha\sum_{i} l_{i}  logf(\theta, X_i)-(1-\alpha)\sum_{i} \hat{l}_ilogf(\theta, X_i)\tag{15}\label{15}\] 
\par
There are two components in loss function:
\begin{enumerate}
    \item first one is hard label loss, which could be any kind of learning-to-rank loss.  Given there are \(n\) items in each training example, Listwise Loss(which is softmax cross entropy) \cite{cao07} is adopted as our hard label loss function, thus \(f(\theta, X_i)\mid_p = \frac{e^{z_{i,p}}}{\sum_{j=1}^n e^{z_{i,j}}}\) where \(z_{i,p}\) is \(X_i\)'s \(p^{th}\) item's score output from model, and \(f(\theta, X_i)\mid_p\) is denoted as softmax prediction of \(p^{th}\) item.   \(l_{i}\) is primary objective's label vector where \(l_{i,p}\) is hard label of \(p^{th}\) item in training example \(X_i\). 
    \item the second component is soft-label distillation loss, it's also a cross-entropy loss following the first distillation paper \cite{Geoffrey14}  , the soft-label \(\hat{l}_i=\sum_{k=1}^{K}\omega_k s_k\) like how it's defined in \eqref{14}.  Similar to \cite{Geoffrey14}, temperature is also applied to softmax inside \(f(\theta, X_i)\).
    \item As \cite{Geoffrey14} proposed, a weighted average is also used to combine hard-label loss and soft-label loss. This is equivalent to construct a new label \(l^{'}_i=\alpha l_{i} + (1-\alpha)\hat{l}_i\) by injecting ground truth label into soft-label.

\end{enumerate}
\par

Figure \ref{fig_1} shows the training graph of multi-objective learning to rank system with model distillation (MO-LTR-MD) based on \eqref{15}. Unless mentioned otherwise all models here are deep learning models, specifically they are all MLP models without advanced model structure:
\begin{figure}[h] 
  \centering
  \includegraphics[width=\linewidth]{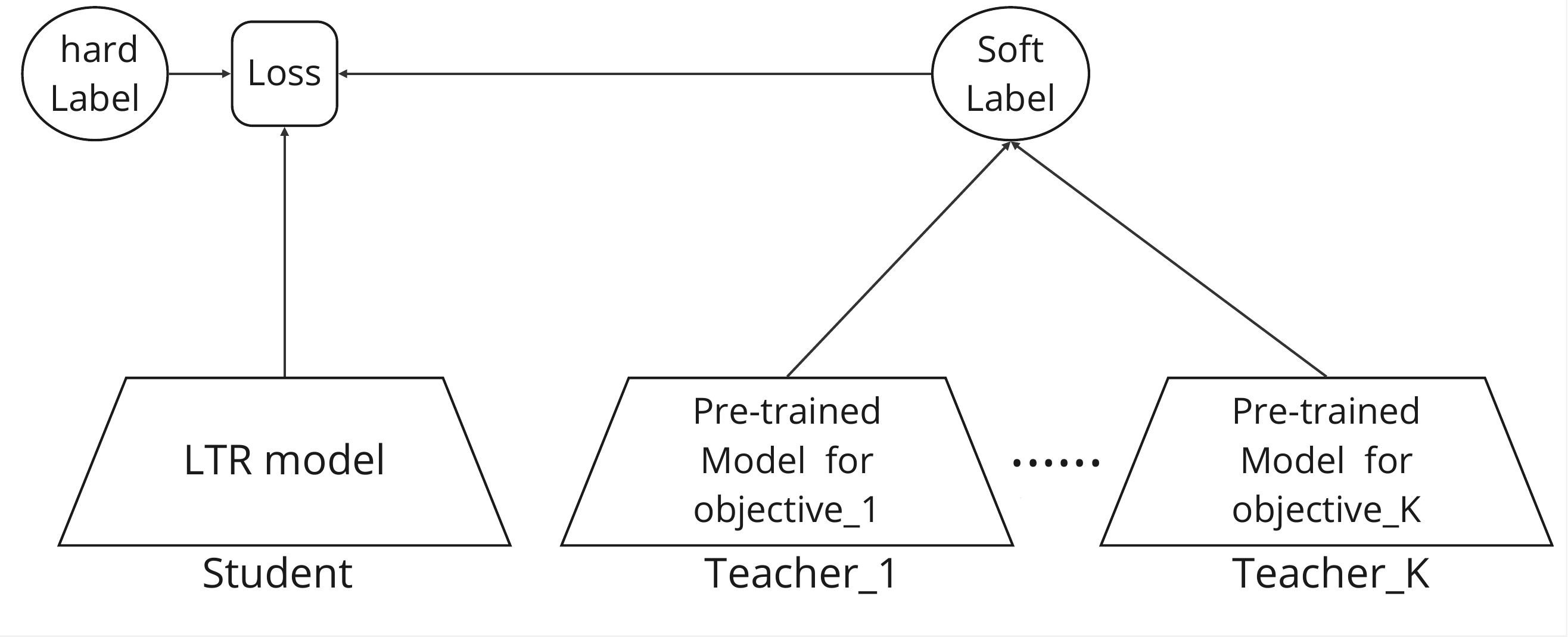}
  \caption{Model training graph}\label{fig_1}
\end{figure}
\begin{itemize}
    \item The hard label is ground truth label of primary objective. It depends on business requirement, usually it could be conversion, click, etc. For most of online marketplaces,  conversion is considered as primary objective.
    \item As mentioned in section \ref{framework} , soft-label is computed from a few pre-trained models:
    \begin{itemize}
        \item  Model is pre-trained for each objective separately, such pre-trained model optimizes each objective cost function independently without considering other objectives, and it's the best effort we could achieve for each objective.
        \item For each training example \(X\), soft label is computed as \(\sum_{k=1}^{K}\omega_k s_k\), where \(s_k=f_k(\theta^*_k, X)\) is ranking score computed from each pre-trained model. 
        \item  During training stage, each pre-trained model is frozen and non-trainable, all the pre-trained models are teacher models in model distillation terminology.
    \end{itemize}
\end{itemize}
\par
During training time,  all models(LTR model and pre-trained models) are loaded into the computation graph, while all pre-trained models' (teacher models') weights are frozen, therefore pre-trained models only contribute to soft-label computation during forward propagation. Later during serving time, all pre-trained models are discard, only the LTR model (student model) is exported for serving.

\subsection{What does soft-label encode?}
Before diving into soft-label, let's recall how hard label (ground truth label) works, in learning to rank scenery hard labels are collected from users' actions, e.g. clicks, purchases, cancellations,returns, etc. Hard labels don't only encode users' preferences but also the ranking model's preference, since only items preferred by model could be shown to users so that labels could be collected from those items. But hard label is highly sparse, e.g. in Airbnb search, each search result page shows a finite site of listings, in most of time booking label is attributed to only one listing among those listings, other labels like cancellations, customer service inquiries could even much sparser. Thus hard label can only help us get partial ground truth : the booked listing is ranked higher than unbooked listings by user, but the oder of unbooked listings is not exposed by users. At another side, soft-label is purely from model score which only encodes model's preference, user preference is missing from soft-label.  With soft-label, full order of all listings could be generated, though this order comes from another model, it still gives new model a good reference (especially when the teacher models are also production models) . The new model could balance knowledge from soft-label and its own hard label during training. \par
In section \ref{softlabel},  soft-label is defined as aggregation of multiple pre-trained models, therefore it actually encodes a simple multi-objective ranking system's preference, this simple system can't be optimal, but there are a few advantages in practice:
\begin{enumerate}
    \item By introducing soft-label, a dense soft-label vector could be assigned to each training example, this could help mitigate imbalanced data issue. While for other multi-objective algorithms like \cite{David20},  when dataset is imbalanced across objectives, they may have to do heavily training data down-sampling or up-sampling, which may hurt model training efficiency.
    \item Soft-label acts as good regularizer(similar to label smoothing regularization\cite{Yuan20}). Soft-label gives multi-objective LTR model feasible prior knowledge since it encodes knowledge from each objective's pre-trained model. It's especially beneficial when the model is bootstrapped from a multi-task learning system \cite{ChunHow23} whose score aggregates each objective model score in the similar way with soft-label.
    \item Soft-label could work well with non-differentiable objective. Besides encoding model preference, as it's shown in section \ref{ad-hoc} soft-label could also encode ad-hoc non-differentiable business objective efficiently.
    \item Soft-label could carry ranking knowledge efficiently and pass to new version of ranking models. In industrial practice, model needs to be regularly retrained so that it could catch new customer trends. Such operation usually takes cold-start approach which trains model from scratch each time, therefore the model totally forgets what it learned before. As section \ref{transfer} shows, by applying soft-label, what model learned before could be passed to new version of model efficiently and also help reduce model irreproducibility and instability.
\end{enumerate}

\subsection{Ad-hoc Business Objective} \label{ad-hoc}
It's very common in practice some objective cost can't be expressed as a differentiable loss function, especially if it comes from ad-hoc business requirements, e.g. show more new items in search results to help new business owner and also marketplace's long term growth, or uprank more higher quality items in top search results to improve marketplace's branding. Such objectives themselves are vague and also can't be optimized by learning from past data, since they are usually in opposite direction to the user behaviors model could observe. For example, users tend to purchase well-established items, model could easily learn this from data and upranks well-established items accordingly, but it's hard to learn that new items need to be upranked. Though it's still possible item-to-item collaborative filtering could help new items, the features(e.g. number of reviews, review ratings, etc) which are highly correlated to some objectives(e.g. CTR, CVR) could still bias the learning heavily towards well-established items, since those features are missing in new items. Thus it's hard for traditional multi-objective optimization to include such ad-hoc non-differentiable objective.\par
One solution could be completing such features for new items by averaging similar well-established items' feature values, but the learning efficiency can't be guaranteed. There is no quantitative way to know the real impact to new items by doing this. And also blindly assuming new items perform at average level may not be fair to items underperformed in past, it may even hurt user experience and other business objectives like CVR. \par
Another solution is to directly revise training data labels. Certainly this can't be done to hard label, e.g. let purchase as 1, non-purchase as 0,  if a small boost \(\beta > 0\) is given to new item's label , while keep other non-purchased items labels as 0, it will end up with purchased item > new items > other items. This doesn't make any sense since it places new items on top of all other items with purchases before. But such boost could work perfectly for soft-labels, e.g. if there is a list of items which are sorted by soft-labels as \(\{t_1, t_2, t_3, t_4, t_5, t_6\}\) in descending order,  where \(t_6\) is new item, now a boost \(\beta\)  is given to score of \(t_6\),  \(t_6\) could be ranked higher than \(t5\) or even \(t_1\).  \(\beta\) could be tuned to control how often and how strong this flip could happen, also since this is applied to soft-labels which have the similar power with hard-label in loss function, the learning could be very efficient. One may argue this boost \(\beta\) could be directly added into ranking score at serving stage, but such change is far more powerful to hurt business metrics, as experiment showed in section \ref{exp:ad-hoc} , it will cause model performance degrade. While adding boost \(\beta\) into soft-label and letting model to balance with other objectives could yield better results.

\subsection{Training Operational Overhead And Irreproducibility } \label{transfer}
In industrial practice, ranking model has to be updated regularly to catch latest consumer trend, users' personalized new preference, etc. Such update could be done daily, weekly, or monthly as offline batch training from scratch(cold start), or even in real-time as online continuous training. One issue of retraining model from scratch is, it will forget what was learned in last version and potentially cause irreproducibility and instability issue. As \cite{Rohan22} pointed out, even same model was trained twice with same data, metrics difference would still be observed between two models. Indeed if our ranking model is trained twice with the same training data, not small difference from side-by-side comparison would be observed. Such irreproducibility would cause more severe instability issue when model is retrained with new data, in general it's expected model trained with new training data would not hurt our metrics at least. But due to the irreproducibility issue, by chance it may end up with a model which performs worse. This is especially important to multi-objective learning to rank, since model is optimized to balance multi-objectives, it's expected that balance among objectives is continuously stable. \par
There may be also some concern of training operational overhead for the proposed system in Figure \ref{fig_1}, since the soft label is computed from pre-trained models, this implies those pre-trained models also have to be retrained regularly(such retraining would also have irreproducibility issue), and the training becomes two-step procedure. 
To address the operational overhead and irreproducibility issues, inspired from Born-Again Neural Networks \cite{Tommaso18} and \cite{Zhen21}, we pass soft-labels down to new version of model during each model retraining in self-distillation way.\par
\begin{figure}[h] 
  \centering
  \includegraphics[width=\linewidth]{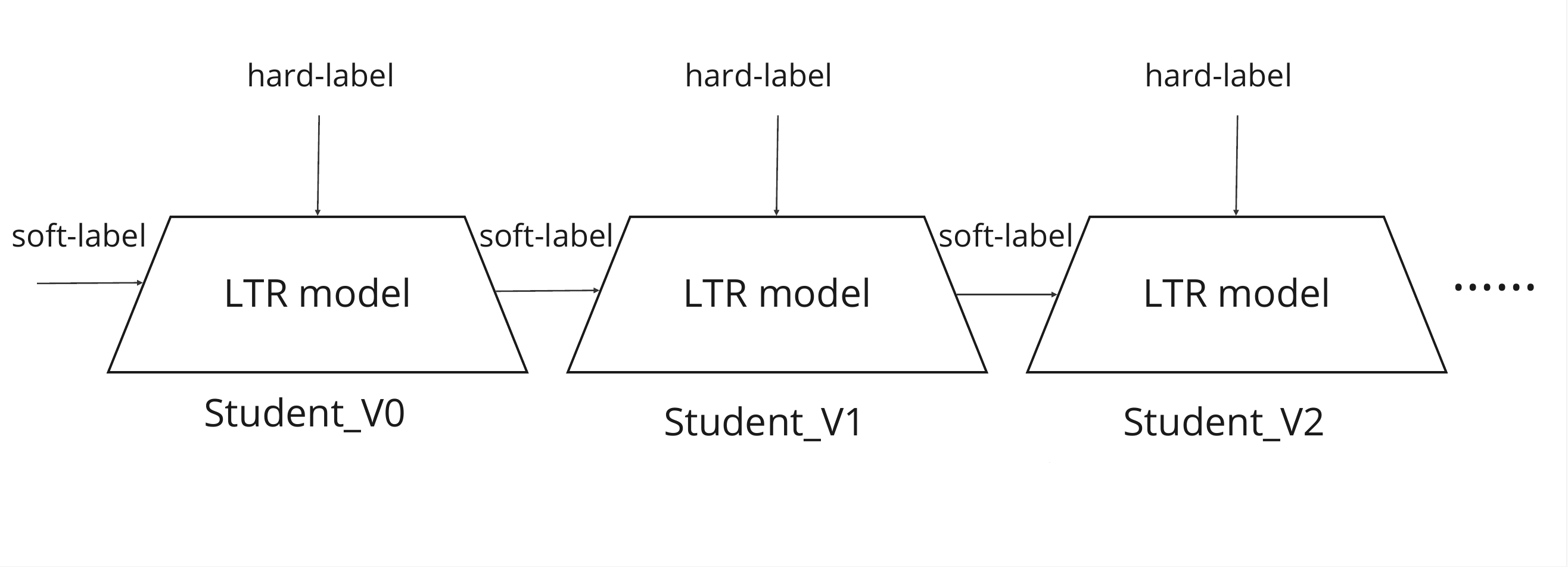}
  \caption{Transfer knowledge among models by soft-label}\label{fig_2}
\end{figure}
Figure \ref{fig_2} shows how it works:
\begin{enumerate}
    \item First version of student model \(V_0\) is trained as Figure \ref{fig_1} describes, it distillates soft-label which is computed from model fusion of each objective's pre-trained model.  
    \item Then to retrain model \(V_1\) over latest training date range, model \(V_1\) is trained from hard label, and also soft-label from ranking score of student model \(V_0\),  instead of the soft-label from model fusion (Note: there is no model structure change from \(V_0\) to \(V_1\)). 
    \item Thus after student model \(V_1\) , the model training is decoupled from pre-trained models, it keeps distilling itself to transfer knowledge learned from each objective's pre-trained model down to new versions.
\end{enumerate}
\par
With this design, the training complexity is significantly reduced: we don't need to maintain and update pre-trained models, instead we distill such knowledge into soft-labels and pass soft-labels along the model training path. But certainly, if needed, those pre-trained models could be plugged-in back any time. Some experiments were also done in section  \ref{exp:self}  and  \ref{exp:reproducibility} to show self-distillation has no negative impact to business metrics and also this innovative approach could help reduce model training irreproducibility and instability. \par

\section{Experimental Results} \label{exp}
\subsection{Experiment Setup}
In this Airbnb experiment, the search ranking system is a multi-objective learning to rank system, the primary objective is booking(CVR), while secondary objectives include  user cancellation rate, host cancellation rate, host rejection rate, platform long term growth, etc.  For booking (CVR), we collect booked listings and \href{https://support.google.com/google-ads/answer/6259715?hl=en}{attribute} \cite{GoogleAttr} them back to searches contain those booked listings, the same attribution process is also done for secondary objectives' labels. \par
The proposed model (MO-LTR-MD) is trained with around 360 millions training examples collected from last a few months which only contains booking label. While each objective's pre-trained model is trained with a shared training dataset contains 500 millions training examples collected in the same date range, each training example in this dataset has multiple labels : bookings, clicks, cancellations, etc. These pre-trained models are co-trained by a multi-task learning system \cite{ChunHow23}. The baseline model is the same multi-task learning system, it's also trained with the same training dataset with 500 millions examples. \par
As mentioned in section \ref{intro}, the baseline model was trained  by combining objectives' loss functions with a set of loss weights, at serving time the ranking score was aggregated from each objective's score by another set of score weights, both weights need to be tuned and tested. While for the proposed MO-LTR-MD model, only one distillation weight \(\alpha\)  in Eq \eqref{15} needs to be tuned, some grid search was done for \(\alpha\), it's found \(\alpha=0.2\) works best for our model. The weights to compute soft-label \eqref{14} came from our production setup which was a model fusion system (we can't share the absolute values of those weights here for protecting our core business data).  Then when we moved to self-distillation stage, the soft-label comes from a score of last version of student model,  the weights in \eqref{14} were not needed any more. \par
\begin{table}
    \centering
    \begin{tabular}{|c|c|c|} \hline 
         Model&  Training Loss weights& Score fusion weights\\ \hline 
         Baseline&  more than K weights& K weights\\ \hline 
         MO-LTR-MD&  one distillation weight& 0\\ \hline
    \end{tabular}
    \caption{Model Training and Serving Comparison}
    \label{tab:training}
\end{table}
Table \ref{tab:training} summarizes the training and serving advantage compared to baseline model:
\begin{enumerate}
    \item Much less training data: 360M V.S. 500M.
    \item Much less loss component weights tuning during training: 1 V.S. K+, where K is number of objectives
    \item No score fusion weights tuning at serving stage: 0 V.S. K
\end{enumerate}
\par

\subsection{Overall model performance}
We evaluate our new system (MO-LTR-MD) with both offline evaluation and online AB test. For offline evaluation, we use Normalized Discountd Cumulative Gain(NDCG) with binary relevance score, where booking is assigned a relevance score of 1 and all other search impressions are assigned a relevance score of 0.\par
The offline evaluation was done over 7 days of data which is not overlapped with our training data, it showed significant improvement of \textbf{+1.1\%} NDCG compared to baseline model which is the multi-task learning system\cite{ChunHow23}.  \par
We also ran an AB test for 3 weeks, the control model is also the multi-task learning system \cite{ChunHow23},  while treatment model is the proposed MO-LTR-MD system. This AB test showed \textbf{+0.37\%} booking(CVR) gain with p\_val = 0.02. As expected, all other secondary objectives' changes were neutral. The AB test showed soft-label does encode knowledge from each objective's pre-trained model, and more importantly because it could act as a regularizer and also mitigate imbalanced data issue, the model could do better job for balancing primary and secondary objectives, such that MO-LTR-MD system could improve NDCG and CVR with such big gain. \par
We also compared the model training cost and serving latency. MO-LTR-MD model was trained with less training data, though pre-trained models were loaded to generate soft-label during training, the total training time was still same compared to baseline model.  Also since only student model was served from MO-LTR-MD system, while multiple ranking models were served in baseline system, the serving latency was significantly reduced : -1.6\% from AB test. 

\subsection{Self-Distillation Test} \label{exp:self}
As we stated in section \ref{transfer}, to make sure multi-objective optimization is stable across model retrainings, and also to reduce operational overhead introduced by model distillation, we extend MO-LTR-MD system as Figure \ref{fig_2} describes. \par
But there are two questions raised for self-distillation : 1) Would self-distillation hurt business metrics given pre-trained models are ablated in this process?  2) If  new training data is added into model training, during self-distillation the pre-trained models are absent and thus can't be updated,  would this hurt model and business metrics?   We designed a test with following steps to address those concerns:
\begin{enumerate}
    \item At first a student model was trained by distilling each objective's pre-trained model as Figure \ref{fig_1} describes,  the generated model is \(V_0\) model in Figure \ref{fig_2} , which still depends on pre-trained models.
    \item   Then as Figure \ref{fig_2} shows another model \(V_1\) was trained with soft-label as ranking score from \(V_0\) model (This is done by doing \(V_0\) model inference over each training example), the training data time window is right shifted a few months .  
    \item As a fair comparison model \(V_0\) was retrained as baseline model,  training data time window was also right shifted by the same number of months.
\end{enumerate}
\par
The offline test shows the NDCG is almost same by comparing \(V_1\) model and retrained \(V_0\) model. This offline test shows transferring ranking knowledge to updated model with only soft-label would not hurt even in absence of pre-trained models, and adding new training data also doesn't dilute soft-label's power. \par

\begin{figure}[h] 
  \centering
  \includegraphics[width=\linewidth]{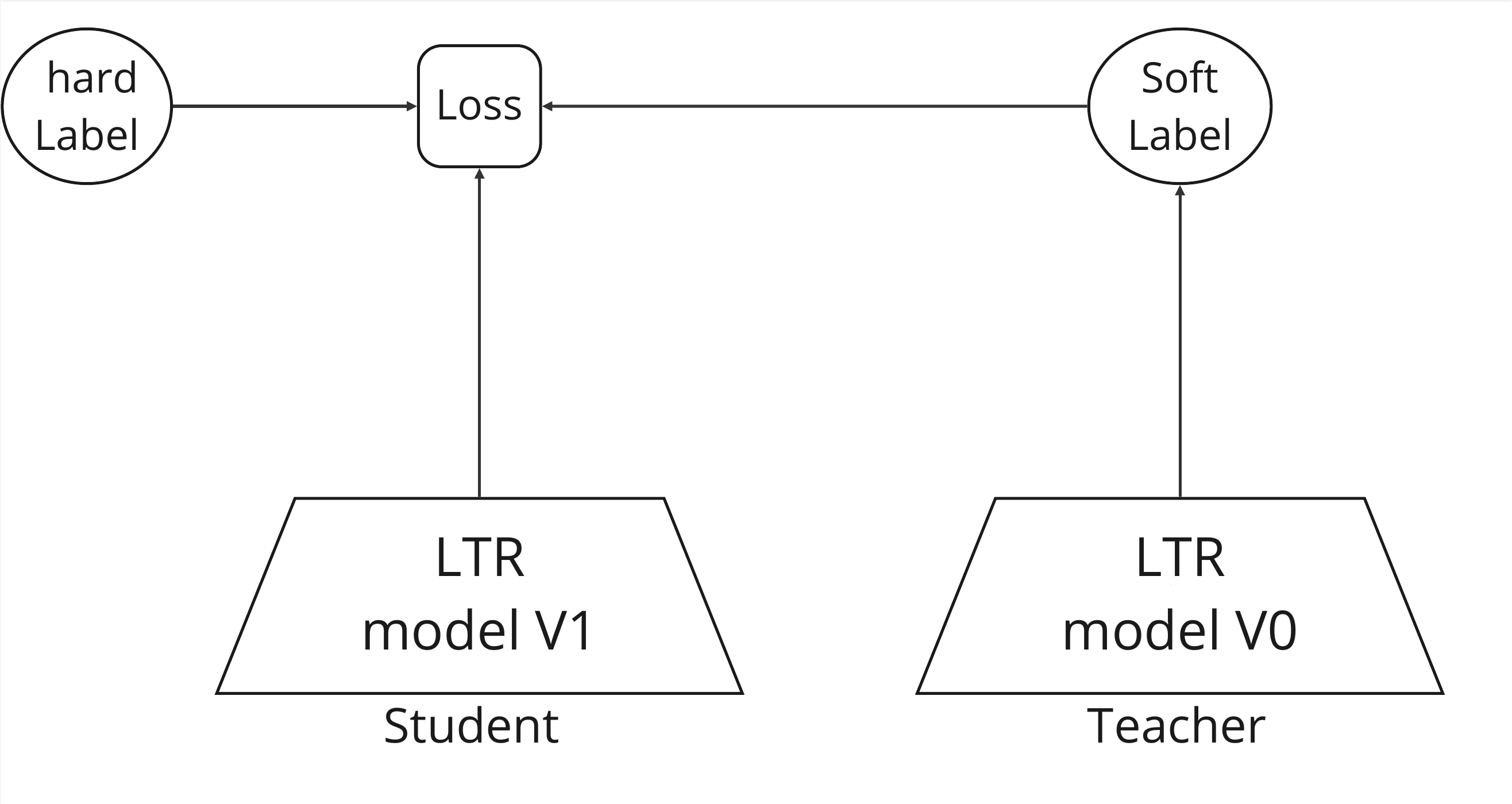}
  \caption{Self distillation}\label{fig_3}
\end{figure}

We also did AB test to verify whether such self-distillation ( knowledge transfer with soft-label) would hurt primary-objective and secondary objectives metrics, the result shows everything is neutral. Thus we could make conclusion that soft-label could efficiently encode and transfer multi-objective optimization knowledge among models. This finding helped us significantly simply the ranking system from Figure \ref{fig_1} to Figure \ref{fig_3}: from now on, model \(V_n\) would be trained from soft-labels computed by model \(V_{n-1}\) and also ground-truth(hard) labels. There is no dependency to pre-trained models any more.\par

\subsection{Model Irreproducibility Test}\label{exp:reproducibility}
To demonstrate soft-label could help reduce irreproducibility, at first we trained a set of same baseline models with same training dataset, baseline models share the same model structure with the student model in MO-LTR-MD system, we only used ground truth label(hard label) to train the baseline models multiple times with different random initial weights. We also trained a set of same MO-LTR-MD models with same training dataset and also different random initial weights. \par
At first we compared models by side-by-side(SxS) comparison system, which looks at two sets of search results, one from the baseline version of ranking model and the other from treatment version we’re testing. The system will review the pages in each set of results, and compute average difference percentage(change rate) between two sets of results. In our system, we applied \href{https://www.statisticshowto.com/kendalls-tau/#:~:text=What%20is%20Kendall's%20Tau%3F,1%20is%20a%20perfect%20relationship.}{Kendall's \(\tau\)} as "a non-parametric measure of relationships between columns of ranked data". Thus we ran SxS among baseline models, and also SxS among MO-LTR-MD models to evaluate model irreproducibility among different retraining. Those SxS could help us understand how search results are flipped by models trained with the same data but different random initial weights. \par
We also computed Relative Prediction Difference(PD) adopted by \cite{Rohan22} to measure irreproducibility, it's defined as 

\[\Delta=1/M \cdot \sum_i \mid \hat{y_{i,1}} - \hat{y_{i,2}} \mid / [(\hat{y_{i,1}} + \hat{y_{i,2}})/2] \]
Where \(\hat{y_{i,1}}\) and \(\hat{y_{i,2}}\)  are predictions from model's different retraining. \par
\begin{table}
    \centering
    \begin{tabular}{|c|c|c|} \hline 
         Model&  Change rate& PD\\ \hline 
         Baseline&  77\%& 0.407\\ \hline 
         MO-LTR-MD&  36\%& 0.363\\ \hline
 Improvement& 53\%&11\%\\\hline
    \end{tabular}
    \caption{Model Irreproducibility}
    \label{tab:reproducibility}
\end{table}
From results in Table \ref{tab:reproducibility}, we could find that with soft-label, SxS change rate(Kenall's \(\tau > 0.02\) ) is reduced by 53\%, PD is reduced by 11\% , both indicate the model irreproducibility is significantly reduced.\par

\subsection{Ad-hoc Business Objective Test} \label{exp:ad-hoc}
Finally we want to test efficiency of injecting ad-hot business objective with soft-label. For protecting our core business data, we can't disclose the real ad-hoc business objective applied in our system.  Instead we simulate some ad-hoc objective which is never applied in our system,  e.g. boosting more high review rating items in top search results. At first we built our baseline by manually giving items with high review rating  a score boost at serving time

\[s_i = \left\{ \begin{array}{rcl}
s_i & \mbox{for}
& r_i < \rho \\ s_i + \alpha & \mbox{for} & r_i \geq \rho 
\end{array}\right. \]
where \(s_i\) is ranking score computed from baseline model at serving time, \(\rho\) is rating threshold, we simply add a boost \(\alpha\) to \(s_i\) if review rating \(r_i\) is better than \(\rho\). \par
Then we built a model by revising soft-label and the model is trained using same data with the baseline model:
\[\hat{l}^{'}_i = \left\{ \begin{array}{rcl}
\hat{l}_i & \mbox{for}
& r_i < \rho \\ \hat{l}_i + \beta & \mbox{for} & r_i \geq \rho 
\end{array}\right. \]
Here we will give soft-label \(\hat{l}_i\) a boost \(\beta\) if review rating \(r_i\) is better than \(\rho\).  During our offline simulations, we carefully tune \(\alpha\) and \(\beta\) to make sure both baseline score boost and our soft-label boost show almost same high rating listings percentage in our SxS test.\par
 
 \begin{table}
     \centering
     \begin{tabular}{|c|c|} \hline 
          Methods&  NDCG\\ \hline 
          directly boost score&  -0.5\%\\ \hline 
          Inject boost to soft-label&  -0.1\%\\ \hline
     \end{tabular}
     \caption{ Boosting Impact To NDCG}
     \label{tab:boosting}
 \end{table}
As Table \ref{tab:boosting}  shows, if we directly give ranking score a boost, we observed \(-0.5\%\) NDCG loss compared to baseline model without score boosting. While if we instead boost soft-labels at training time, to get the same impact to high rating listings, we only have to sacrifice \(-0.1\%\) NDCG loss compared to the same baseline model.  Why does soft-label injection hurt metric much less ? Deep learning model training usually is a non-convex optimization problem, thus it may end up with different local minimums. As mentioned before soft-label could be considered as a regularizer, by injecting some boost into regularizer, we actually pushed model towards the local minimum more aligned with ad-hoc business objective.  \par

\section{Learning and Future Work}
Our experiments demonstrated the proposed MO-LTR-MD system doesn't only help us find better optimization solution and improve model irrproducibility, but also shows advantage to work with ad-hoc non-differentiable objective. One interesting learning we had is that during our self-distillation test, the multi-objective optimization ranking knowledge encoded by soft-label could be transferred to new model in stable way by self-distillation. This is verified by our offline simulation and also online test, though at online test we only tested a model with two rounds of self-distillation, we would do more follow up online tests to make sure the knowledge transfer would not be decayed in long run, especially most of metrics for secondary objectives can only be checked in online test. \par
\begin{figure}[h] 
  \centering
  \includegraphics[width=\linewidth]{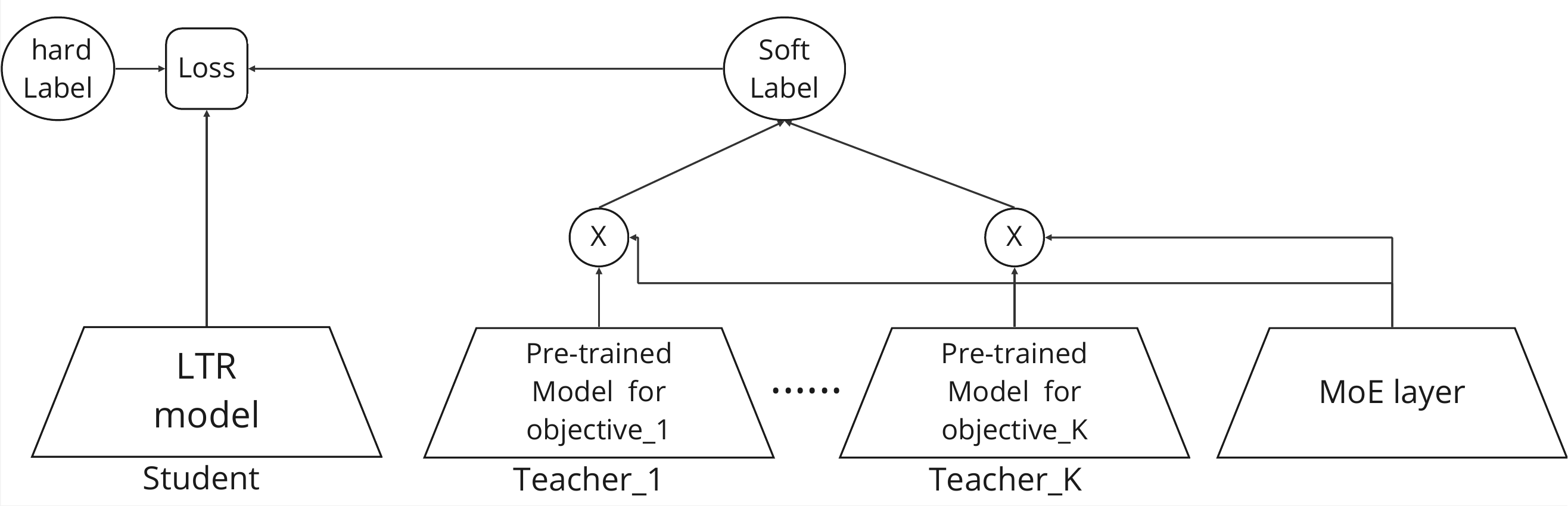}
  \caption{Multi-objective LTR with MoE layer}\label{fig_4}
\end{figure}

In our experiment setup, we bootstrap our distillation based system from existing production model fusion setup so we don't need to tune weights of soft-labels in Eq \eqref{14}. While if we are building a multi-objective learning to rank system from scratch, those weights need to be set, as Figure \ref{fig_4} shows we could fold weights learning into model training by importing a MoE layer \cite{Yulong20}, this work could be done in future when we redesign our multi-objective ranking system.
 
\section{Acknowledgments}
We'd like to thank Do-kyum Kim for inspiring us to start this work, and also lots of good discussion along the whole path, and also Chunhow Tan, Han Zhao for the discussion and support. We would also like to extend our thanks to Alex Deng who builds Airbnb interleaving system which enables agile model development and online test, also to Pavan Tapadia and Sherry Chen for product support. Finally, we would like to thank the entire Airbnb Relevance team for valuable discussions.

\bibliographystyle{ACM-Reference-Format}
\bibliography{sample-base}


\begin{thebibliography}{29}


\ifx \showCODEN    \undefined \def \showCODEN     #1{\unskip}     \fi
\ifx \showDOI      \undefined \def \showDOI       #1{#1}\fi
\ifx \showISBNx    \undefined \def \showISBNx     #1{\unskip}     \fi
\ifx \showISBNxiii \undefined \def \showISBNxiii  #1{\unskip}     \fi
\ifx \showISSN     \undefined \def \showISSN      #1{\unskip}     \fi
\ifx \showLCCN     \undefined \def \showLCCN      #1{\unskip}     \fi
\ifx \shownote     \undefined \def \shownote      #1{#1}          \fi
\ifx \showarticletitle \undefined \def \showarticletitle #1{#1}   \fi
\ifx \showURL      \undefined \def \showURL       {\relax}        \fi
\providecommand\bibfield[2]{#2}
\providecommand\bibinfo[2]{#2}
\providecommand\natexlab[1]{#1}
\providecommand\showeprint[2][]{arXiv:#2}

\bibitem[Goo({[n.\,d.]})]%
        {GoogleAttr}
 \bibinfo{year}{[n.\,d.]}\natexlab{}.
\newblock \bibinfo{booktitle}{\emph{About attribution models}}.
\newblock
\urldef\tempurl%
\url{https://support.google.com/google-ads/answer/6259715?hl=en}
\showURL{%
\tempurl}


\bibitem[Anil et~al\mbox{.}(2022)]%
        {Rohan22}
\bibfield{author}{\bibinfo{person}{Rohan Anil}, \bibinfo{person}{Sandra Gadanho}, \bibinfo{person}{Da Huang}, {and} \bibinfo{person}{Nijith~Jacob etc}.} \bibinfo{year}{2022}\natexlab{}.
\newblock \showarticletitle{On the Factory Floor: ML Engineering for Industrial-Scale Ads Recommendation Models}. In \bibinfo{booktitle}{\emph{Recsys 2022 Workshop on Online Recommender Systems and User Modeling}}. \bibinfo{publisher}{ACM}, \bibinfo{address}{New York, NY}.
\newblock


\bibitem[Burges et~al\mbox{.}(2005)]%
        {Burges05}
\bibfield{author}{\bibinfo{person}{C. Burges}, \bibinfo{person}{T. Shaked}, \bibinfo{person}{E. Renshaw}, \bibinfo{person}{A. Lazier}, \bibinfo{person}{M. Deeds}, \bibinfo{person}{N. Hamilton}, {and} \bibinfo{person}{G. Hullender}.} \bibinfo{year}{2005}\natexlab{}.
\newblock \showarticletitle{Learning to rank using gradient descent}. In \bibinfo{booktitle}{\emph{ICML ’05: Proceedings of the 22nd International Conference on Machine learning}}. \bibinfo{pages}{89--96}.
\newblock


\bibitem[Burges(2010)]%
        {burges2010from}
\bibfield{author}{\bibinfo{person}{Chris~J.C. Burges}.} \bibinfo{year}{2010}\natexlab{}.
\newblock \bibinfo{booktitle}{\emph{From RankNet to LambdaRank to LambdaMART: An Overview}}.
\newblock \bibinfo{type}{{T}echnical {R}eport} MSR-TR-2010-82.
\newblock


\bibitem[Cao et~al\mbox{.}(2007)]%
        {cao07}
\bibfield{author}{\bibinfo{person}{Z. Cao}, \bibinfo{person}{T. Qin}, \bibinfo{person}{T.-Y. Liu}, \bibinfo{person}{M.-F. Tsai}, \bibinfo{person}{}, {and} \bibinfo{person}{H. Li}.} \bibinfo{year}{2007}\natexlab{}.
\newblock \showarticletitle{Learning to rank: from pairwise approach to listwise approach}. In \bibinfo{booktitle}{\emph{ICML ’07: Proceedings of the 24th International Conference on Machine learning}}. \bibinfo{pages}{129--136}.
\newblock


\bibitem[Carmel et~al\mbox{.}(2020)]%
        {David20}
\bibfield{author}{\bibinfo{person}{David Carmel}, \bibinfo{person}{Elad Haramaty}, \bibinfo{person}{Arnon Lazerson}, {and} \bibinfo{person}{Liane Lewin-Eytan}.} \bibinfo{year}{2020}\natexlab{}.
\newblock \showarticletitle{Multi-Objective Ranking Optimization for Product Search Using Stochastic Label Aggregation}. In \bibinfo{booktitle}{\emph{WWW '20: Proceedings of The Web Conference 2020}}. \bibinfo{publisher}{ACM}, \bibinfo{address}{New York, NY}, \bibinfo{pages}{373--383}.
\newblock


\bibitem[Furlanello et~al\mbox{.}(2018)]%
        {Tommaso18}
\bibfield{author}{\bibinfo{person}{Tommaso Furlanello}, \bibinfo{person}{Zachary~C. Lipton}, \bibinfo{person}{Michael Tschannen}, \bibinfo{person}{Laurent Itti}, {and} \bibinfo{person}{Anima Anandkumar}.} \bibinfo{year}{2018}\natexlab{}.
\newblock \showarticletitle{Born-Again Neural Networks}. In \bibinfo{booktitle}{\emph{ICML2018: Proceedings of the 35 th International Conference on Machine Learning}}.
\newblock


\bibitem[Geoffrey~Hinton(2014)]%
        {Geoffrey14}
\bibfield{author}{\bibinfo{person}{Jeff~Dean Geoffrey~Hinton, Oriol~Vinyals}.} \bibinfo{year}{2014}\natexlab{}.
\newblock \showarticletitle{Distilling the Knowledge in a Neural Network}. In \bibinfo{booktitle}{\emph{NIPS 2014 Deep Learning Workshop}}.
\newblock


\bibitem[Gu et~al\mbox{.}(2020)]%
        {Yulong20}
\bibfield{author}{\bibinfo{person}{Yulong Gu}, \bibinfo{person}{Zhuoye Ding}, {and} \bibinfo{person}{Shuaiqiang~Wang etc}.} \bibinfo{year}{2020}\natexlab{}.
\newblock \showarticletitle{Deep Multifaceted Transformers for Multi-objective Ranking in Large-Scale E-commerce Recommender Systems}. In \bibinfo{booktitle}{\emph{CIKM '20: Proceedings of the 29th ACM International Conference on Information and Knowledge Management}}. \bibinfo{publisher}{ACM}, \bibinfo{address}{New York, NY}, \bibinfo{pages}{2493--2500}.
\newblock


\bibitem[Hadash et~al\mbox{.}(2018)]%
        {Guy18}
\bibfield{author}{\bibinfo{person}{Guy Hadash}, \bibinfo{person}{Oren~Sar Shalom}, {and} \bibinfo{person}{Rita Osadchy}.} \bibinfo{year}{2018}\natexlab{}.
\newblock \showarticletitle{Rank and rate: multi-task learning for recommender systems}. In \bibinfo{booktitle}{\emph{RecSys '18: Proceedings of the 12th ACM Conference on Recommender Systems}}. \bibinfo{publisher}{ACM}, \bibinfo{address}{New York, NY}, \bibinfo{pages}{451--454}.
\newblock


\bibitem[Lin et~al\mbox{.}(2019)]%
        {Xiao19}
\bibfield{author}{\bibinfo{person}{Xiao Lin}, \bibinfo{person}{Hongjie Chen}, {and} \bibinfo{person}{Changhua~Pei etc}.} \bibinfo{year}{2019}\natexlab{}.
\newblock \showarticletitle{A Pareto-Eficient Algorithm for Multiple Objective Optimization in E-Commerce Recommendation}. In \bibinfo{booktitle}{\emph{RecSys '19: Proceedings of the 13th ACM Conference on Recommender Systems}}. \bibinfo{publisher}{ACM}, \bibinfo{address}{New York, NY}, \bibinfo{pages}{20--28}.
\newblock


\bibitem[Mahapatra et~al\mbox{.}(2023)]%
        {Debabrata23}
\bibfield{author}{\bibinfo{person}{Debabrata Mahapatra}, \bibinfo{person}{Chaosheng Dong}, \bibinfo{person}{Yetian Chen}, {and} \bibinfo{person}{Michinari Momma}.} \bibinfo{year}{2023}\natexlab{}.
\newblock \showarticletitle{Multi-Label Learning to Rank through Multi-Objective Optimization}. In \bibinfo{booktitle}{\emph{KDD '23: Proceedings of the 29th ACM SIGKDD Conference on Knowledge Discovery and Data Mining}}. \bibinfo{publisher}{ACM}, \bibinfo{address}{New York, NY}, \bibinfo{pages}{4605--4616}.
\newblock


\bibitem[McMahan et~al\mbox{.}(2013)]%
        {McMahan13}
\bibfield{author}{\bibinfo{person}{H.~Brendan McMahan}, \bibinfo{person}{Gary Holt}, \bibinfo{person}{D.Sculley}, {and} \bibinfo{person}{Michael~Young etc}.} \bibinfo{year}{2013}\natexlab{}.
\newblock \showarticletitle{Ad Click Prediction: a View from the Trenches}. In \bibinfo{booktitle}{\emph{KDD '13: Proceedings of the 19th ACM SIGKDD international conference on Knowledge discovery and data mining}}. \bibinfo{publisher}{ACM}, \bibinfo{address}{New York, NY}, \bibinfo{pages}{1222--1230}.
\newblock


\bibitem[Miettinen(1998)]%
        {Kaisa98}
\bibfield{author}{\bibinfo{person}{Kaisa Miettinen}.} \bibinfo{year}{1998}\natexlab{}.
\newblock \bibinfo{booktitle}{\emph{Nonlinear multiobjective optimization}}. \bibinfo{series}{International Series in Operations Research and Management Science}, Vol.~\bibinfo{volume}{12}.
\newblock \bibinfo{publisher}{Springer}, \bibinfo{address}{New York, NY}.
\newblock


\bibitem[Momma et~al\mbox{.}(2019)]%
        {Michinari19}
\bibfield{author}{\bibinfo{person}{Michinari Momma}, \bibinfo{person}{Alireza~Bagheri Garakani}, {and} \bibinfo{person}{Yi Sun}.} \bibinfo{year}{2019}\natexlab{}.
\newblock \showarticletitle{Multi-objective Relevance Ranking}. In \bibinfo{booktitle}{\emph{Proceedings of ACM SIGIR Workshop on eCommerce (SIGIR 2019 eCom)}}. \bibinfo{publisher}{ACM}, \bibinfo{address}{New York, NY}.
\newblock


\bibitem[Nardini et~al\mbox{.}(2023)]%
        {Franco23}
\bibfield{author}{\bibinfo{person}{Franco~Maria Nardini}, \bibinfo{person}{Cosimo Rulli}, \bibinfo{person}{Salvatore Trani}, {and} \bibinfo{person}{Rossano Venturini}.} \bibinfo{year}{2023}\natexlab{}.
\newblock \showarticletitle{Distilled Neural Networks for Efficient Learning to Rank}.
\newblock \bibinfo{journal}{\emph{IEEE Transactions on Knowledge and Data Engineering}}  \bibinfo{volume}{35} (\bibinfo{date}{May} \bibinfo{year}{2023}), \bibinfo{pages}{4695--4712}.
\newblock
\urldef\tempurl%
\url{https://doi.org/10.1109/TKDE.2022.3152585}
\showDOI{\tempurl}


\bibitem[Qin et~al\mbox{.}(2021)]%
        {Zhen21}
\bibfield{author}{\bibinfo{person}{Zhen Qin}, \bibinfo{person}{Le Yan}, \bibinfo{person}{Yi Tay}, \bibinfo{person}{Honglei Zhuang}, \bibinfo{person}{Xuanhui Wang}, \bibinfo{person}{Michael Bendersky}, \bibinfo{person}{}, {and} \bibinfo{person}{Marc Najork}.} \bibinfo{year}{2021}\natexlab{}.
\newblock \bibinfo{title}{Born again neural rankers}.
\newblock
\newblock
\showeprint[arxiv]{2109.15285}


\bibitem[Ribeiro et~al\mbox{.}(2012)]%
        {Marco12}
\bibfield{author}{\bibinfo{person}{Marco~Tulio Ribeiro}, \bibinfo{person}{Anisio Lacerda}, \bibinfo{person}{Adriano Veloso}, {and} \bibinfo{person}{Nivio Ziviani}.} \bibinfo{year}{2012}\natexlab{}.
\newblock \showarticletitle{Pareto-efficient hybridization for multi-objective recommender systems}. In \bibinfo{booktitle}{\emph{RecSys '12: Proceedings of the sixth ACM conference on Recommender systems}}. \bibinfo{publisher}{ACM}, \bibinfo{address}{New York, NY}, \bibinfo{pages}{19--26}.
\newblock


\bibitem[Rodriguez et~al\mbox{.}(2012)]%
        {Mario12}
\bibfield{author}{\bibinfo{person}{Mario Rodriguez}, \bibinfo{person}{Christian Posse}, {and} \bibinfo{person}{Ethan Zhang}.} \bibinfo{year}{2012}\natexlab{}.
\newblock \showarticletitle{Multiple objective optimization in recommender systems}. In \bibinfo{booktitle}{\emph{RecSys '12: Proceedings of the sixth ACM conference on Recommender systems}}. \bibinfo{publisher}{ACM}, \bibinfo{address}{New York, NY}, \bibinfo{pages}{11--18}.
\newblock


\bibitem[Sener and Koltun(2018)]%
        {Ozan18}
\bibfield{author}{\bibinfo{person}{Ozan Sener} {and} \bibinfo{person}{Vladlen Koltun}.} \bibinfo{year}{2018}\natexlab{}.
\newblock \showarticletitle{Multi-Task Learning as Multi-Objective Optimization}. In \bibinfo{booktitle}{\emph{NIPS'18: Proceedings of the 32nd International Conference on Neural Information Processing Systems}}. \bibinfo{pages}{525--536}.
\newblock


\bibitem[Tan et~al\mbox{.}(2023)]%
        {ChunHow23}
\bibfield{author}{\bibinfo{person}{ChunHow Tan}, \bibinfo{person}{Austin Chan}, \bibinfo{person}{Malay Haldar}, {and} \bibinfo{person}{Jie~Tang etc}.} \bibinfo{year}{2023}\natexlab{}.
\newblock \showarticletitle{Optimizing Airbnb Search Journey with Multi-task Learning}. In \bibinfo{booktitle}{\emph{KDD '23: Proceedings of the 29th ACM SIGKDD Conference on Knowledge Discovery and Data Mining}}. \bibinfo{publisher}{ACM}, \bibinfo{address}{New York, NY}, \bibinfo{pages}{4872--4881}.
\newblock


\bibitem[Tang and Wang(2018)]%
        {Jiaxi18}
\bibfield{author}{\bibinfo{person}{Jiaxi Tang} {and} \bibinfo{person}{Ke Wang}.} \bibinfo{year}{2018}\natexlab{}.
\newblock \showarticletitle{Ranking Distillation: Learning Compact Ranking Models With High Performance for Recommender System}. In \bibinfo{booktitle}{\emph{KDD '18: Proceedings of the 24th ACM SIGKDD International Conference on Knowledge Discovery and Data Mining}}. \bibinfo{publisher}{ACM}, \bibinfo{address}{New York, NY}, \bibinfo{pages}{2289--2298}.
\newblock


\bibitem[Woo and amd Sangwoo~Cho(2022)]%
        {Honguk22}
\bibfield{author}{\bibinfo{person}{Honguk Woo} {and} \bibinfo{person}{Hyunsung~Lee amd Sangwoo~Cho}.} \bibinfo{year}{2022}\natexlab{}.
\newblock \showarticletitle{An Efficient Combinatorial Optimization Model Using Learning-to-Rank Distillation}. In \bibinfo{booktitle}{\emph{The Thirty-Sixth AAAI Conference on Artificial Intelligence (AAAI-22)}}. \bibinfo{pages}{8666--8674}.
\newblock


\bibitem[Xie et~al\mbox{.}(2021)]%
        {Ruobing21}
\bibfield{author}{\bibinfo{person}{Ruobing Xie}, \bibinfo{person}{Yanlei Liu}, {and} \bibinfo{person}{Shaoliang~Zhang etc}.} \bibinfo{year}{2021}\natexlab{}.
\newblock \showarticletitle{Personalized Approximate Pareto-Efficient Recommendation}. In \bibinfo{booktitle}{\emph{WWW '21: Proceedings of the Web Conference 2021}}. \bibinfo{publisher}{ACM}, \bibinfo{address}{New York, NY}, \bibinfo{pages}{3839--3849}.
\newblock


\bibitem[Yuan et~al\mbox{.}(2020)]%
        {Yuan20}
\bibfield{author}{\bibinfo{person}{Li Yuan}, \bibinfo{person}{Francis~EH Tay}, {and} \bibinfo{person}{Guilin~Li etc}.} \bibinfo{year}{2020}\natexlab{}.
\newblock \showarticletitle{Revisiting Knowledge Distillation via Label Smoothing Regularization}. In \bibinfo{booktitle}{\emph{2020 IEEE/CVF Conference on Computer Vision and Pattern Recognition (CVPR)}}.
\newblock


\bibitem[Zhang et~al\mbox{.}(2021)]%
        {Honglei21}
\bibfield{author}{\bibinfo{person}{Honglei Zhang}, \bibinfo{person}{Zhen Qin}, \bibinfo{person}{Shuguang Han}, {and} \bibinfo{person}{Xuanhui~Wang etc}.} \bibinfo{year}{2021}\natexlab{}.
\newblock \showarticletitle{Ensemble Distillation for BERT-Based Ranking Models}. In \bibinfo{booktitle}{\emph{ICTIR '21: Proceedings of the 2021 ACM SIGIR International Conference on Theory of Information Retrieval}}. \bibinfo{publisher}{ACM}, \bibinfo{address}{New York, NY}, \bibinfo{pages}{131--136}.
\newblock


\bibitem[Zhao et~al\mbox{.}(2023)]%
        {Zhishan23}
\bibfield{author}{\bibinfo{person}{Zhishan Zhao}, \bibinfo{person}{Jingyue Gao}, {and} \bibinfo{person}{Yu~Zhang etc}.} \bibinfo{year}{2023}\natexlab{}.
\newblock \showarticletitle{COPR: Consistency-Oriented Pre-Ranking for Online Advertising}. In \bibinfo{booktitle}{\emph{CIKM '23: Proceedings of the 32nd ACM International Conference on Information and Knowledge Management}}. \bibinfo{publisher}{ACM}, \bibinfo{address}{New York, NY}, \bibinfo{pages}{4974–4980}.
\newblock


\bibitem[Zhao et~al\mbox{.}(2019)]%
        {Zhe18}
\bibfield{author}{\bibinfo{person}{Zhe Zhao}, \bibinfo{person}{Lichan Hong}, {and} \bibinfo{person}{Li~Wei etc}.} \bibinfo{year}{2019}\natexlab{}.
\newblock \showarticletitle{Recommending what video to watch next: a multitask ranking system}. In \bibinfo{booktitle}{\emph{RecSys '19: Proceedings of the 13th ACM Conference on Recommender Systems}}. \bibinfo{publisher}{ACM}, \bibinfo{address}{New York, NY}, \bibinfo{pages}{43--51}.
\newblock


\bibitem[Zheng and Wang(2022)]%
        {Yong22}
\bibfield{author}{\bibinfo{person}{Yong Zheng} {and} \bibinfo{person}{David~(Xuejun) Wang}.} \bibinfo{year}{2022}\natexlab{}.
\newblock \showarticletitle{A survey of recommender systems with multi-objective optimization}.
\newblock \bibinfo{journal}{\emph{Neurocomputing}}  \bibinfo{volume}{474} (\bibinfo{date}{Feb.} \bibinfo{year}{2022}), \bibinfo{pages}{141--153}.
\newblock
\urldef\tempurl%
\url{https://doi.org/10.1016/j.neucom.2021.11.041}
\showDOI{\tempurl}


\end{thebibliography}

\section{Appendices}
\subsection{A.1} \label{proof}
Short proof that we could simplify \eqref{11} to \eqref{12}

\begin{align*}
& \sum_{k=1}^{K} \omega_k CE(f(\theta,X), f_k(\theta^*_k,X)) \\
& = \sum_{k=1}^{K} \omega_k \sum_{i}-f_k(\theta^*_k, X_i) \cdot logf(\theta, X_i) \\
& =\sum_{i} -logf(\theta, X_i) \cdot \sum_{k=1}^{K}\omega_k f_k(\theta^*_k, X_i) \\
& =CE(f(\theta,X), \sum_{k=1}^{K}\omega_k f_k(\theta^*_k,X))
\end{align*}

\end{document}